\documentclass[twoside,twocolumn,9pt]{article}
\usepackage{extsizes}
\usepackage[super,sort&compress,comma]{natbib} 
\usepackage[version=3]{mhchem}
\usepackage[left=1.5cm, right=1.5cm, top=1.785cm, bottom=2.0cm]{geometry}
\usepackage{balance}
\usepackage{mathptmx}
\usepackage{sectsty}
\usepackage{graphicx} 
\usepackage{lastpage}
\usepackage[format=plain,justification=justified,singlelinecheck=false,font={stretch=1.125,small,sf},labelfont=bf,labelsep=space]{caption}
\usepackage{float}
\usepackage{fancyhdr}
\usepackage{fnpos}
\usepackage[english]{babel}
\addto{\captionsenglish}{%
  
}
\usepackage{array}
\usepackage{droidsans}
\usepackage{charter}
\usepackage[T1]{fontenc}
\usepackage[usenames,dvipsnames]{xcolor}
\usepackage{setspace}
\usepackage[compact]{titlesec}
\usepackage{hyperref}

\usepackage{epstopdf}

\usepackage{amssymb}
\usepackage{bm}

\newcommand{\inner}[2]{{\left\langle{#1},{#2}\right\rangle}}

\usepackage{color}

\definecolor{cream}{RGB}{222,217,201}
\begin{document}

\pagestyle{fancy}
\thispagestyle{plain}
\fancypagestyle{plain}{
\renewcommand{\headrulewidth}{0pt}
}

\makeFNbottom
\makeatletter
\renewcommand\LARGE{\@setfontsize\LARGE{15pt}{17}}
\renewcommand\Large{\@setfontsize\Large{12pt}{14}}
\renewcommand\large{\@setfontsize\large{10pt}{12}}
\renewcommand\footnotesize{\@setfontsize\footnotesize{7pt}{10}}
\renewcommand\scriptsize{\@setfontsize\scriptsize{7pt}{7}}
\makeatother

\renewcommand{\thefootnote}{\fnsymbol{footnote}}
\renewcommand\footnoterule{\vspace*{1pt}%
\color{cream}\hrule width 3.5in height 0.4pt \color{black} \vspace*{5pt}} 
\setcounter{secnumdepth}{5}

\makeatletter 
\renewcommand\@biblabel[1]{#1}            
\renewcommand\@makefntext[1]%
{\noindent\makebox[0pt][r]{\@thefnmark\,}#1}
\makeatother 
\renewcommand{\figurename}{\small{Fig.}~}
\sectionfont{\sffamily\Large}
\subsectionfont{\normalsize}
\subsubsectionfont{\bf}
\setstretch{1.125} 
\setlength{\skip\footins}{0.8cm}
\setlength{\footnotesep}{0.25cm}
\setlength{\jot}{10pt}
\titlespacing*{\section}{0pt}{4pt}{4pt}
\titlespacing*{\subsection}{0pt}{15pt}{1pt}

\fancyfoot{}
\fancyfoot[LO,RE]{\vspace{-7.1pt}\includegraphics[height=9pt]{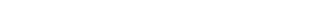}}
\fancyfoot[CO]{\vspace{-7.1pt}\hspace{13.2cm}\includegraphics{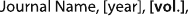}}
\fancyfoot[CE]{\vspace{-7.2pt}\hspace{-14.2cm}\includegraphics{head_foot/RF}}
\fancyfoot[RO]{\footnotesize{\sffamily{1--\pageref{LastPage} ~\textbar  \hspace{2pt}\thepage}}}
\fancyfoot[LE]{\footnotesize{\sffamily{\thepage~\textbar\hspace{3.45cm} 1--\pageref{LastPage}}}}
\fancyhead{}
\renewcommand{\headrulewidth}{0pt} 
\renewcommand{\footrulewidth}{0pt}
\setlength{\arrayrulewidth}{1pt}
\setlength{\columnsep}{6.5mm}
\setlength\bibsep{1pt}

\makeatletter 
\newlength{\figrulesep} 
\setlength{\figrulesep}{0.5\textfloatsep} 

\newcommand{\topfigrule}{\vspace*{-1pt}%
\noindent{\color{cream}\rule[-\figrulesep]{\columnwidth}{1.5pt}} }

\newcommand{\botfigrule}{\vspace*{-2pt}%
\noindent{\color{cream}\rule[\figrulesep]{\columnwidth}{1.5pt}} }

\newcommand{\dblfigrule}{\vspace*{-1pt}%
\noindent{\color{cream}\rule[-\figrulesep]{\textwidth}{1.5pt}} }

\makeatother

\twocolumn[
  \begin{@twocolumnfalse}
{\includegraphics[height=30pt]{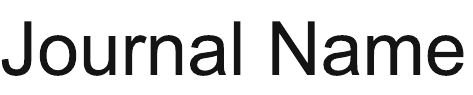}\hfill\raisebox{0pt}[0pt][0pt]{\includegraphics[height=55pt]{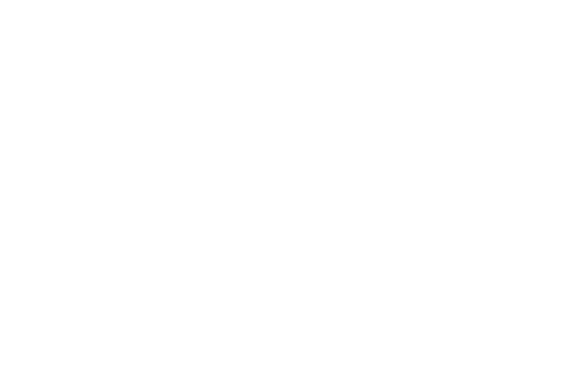}}\\[1ex]
\includegraphics[width=18.5cm]{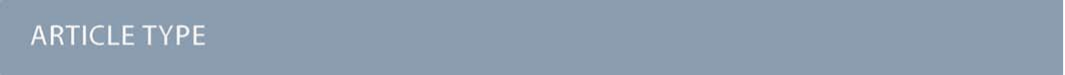}}\par
\vspace{1em}
\sffamily
\begin{tabular}{m{4.5cm} p{13.5cm} }

\includegraphics{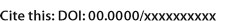} & \noindent\LARGE{\textbf{From Polyhedra to Crystals: A Graph-Theoretic Framework for Crystal Structure Generation$^\dag$}} \\
 & \vspace{0.3cm} \\

 & \noindent\large{Tomoyasu Yokoyama,$^{\ast}$\textit{$^{a}$} Kazuhide Ichikawa,\textit{$^{a}$} and Hisashi Naito\textit{$^{b}$}} \\

\includegraphics{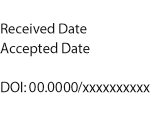} & \\

\end{tabular}

 \end{@twocolumnfalse} \vspace{0.6cm}

  ]

\renewcommand*\rmdefault{bch}\normalfont\upshape
\rmfamily
\section*{}
\vspace{-1cm}


\footnotetext{\textit{$^{a}$~Green Transformation Division, Panasonic Holdings Corporation, Osaka, Japan. E-mail: yokoyama.tomoyasu@jp.panasonic.com}}
\footnotetext{\textit{$^{b}$~Graduate School of Mathematics, Nagoya University, Nagoya, Japan.}}





\sffamily{Crystal structures can be viewed as assemblies of space-filling polyhedra, which play a critical role in determining material properties such as ionic conductivity and dielectric constant. However, most conventional crystal structure prediction methods rely on random structure generation and do not explicitly incorporate polyhedral tiling, limiting their efficiency and interpretability. In this highlight, we introduced a novel crystal structure generation method based on discrete geometric analysis of polyhedral information. The geometry and topology of space-filling polyhedra are encoded as a dual periodic graph, and the corresponding crystal structure is obtained via the standard realization of this graph. We demonstrate the effectiveness of our approach by reconstructing face-centered cubic (FCC), hexagonal close-packed (HCP), and body-centered cubic (BCC) structures from their dual periodic graphs. This method offers a new pathway for systematically generating crystal structures based on target polyhedra, potentially accelerating the discovery of novel materials for applications in electronics, energy storage, and beyond.}\\


\rmfamily 


\section{Introduction}
The essential parameters that determine a material properties are its composition and structure. Composition-based design—such as element substitution and alloying—has achieved significant success. In contrast, crystal structure design remains far more challenging. Compared to composition, structure-dependent properties such as ionic conductivity, dielectric response, and magnetism are less well understood.
Overcoming this limitation could open a new avenue of materials discovery by enabling structure-based design, which may reveal functional materials beyond the reach of composition design alone.

\begin{figure}[h]
\centering
  \includegraphics[width=9cm]{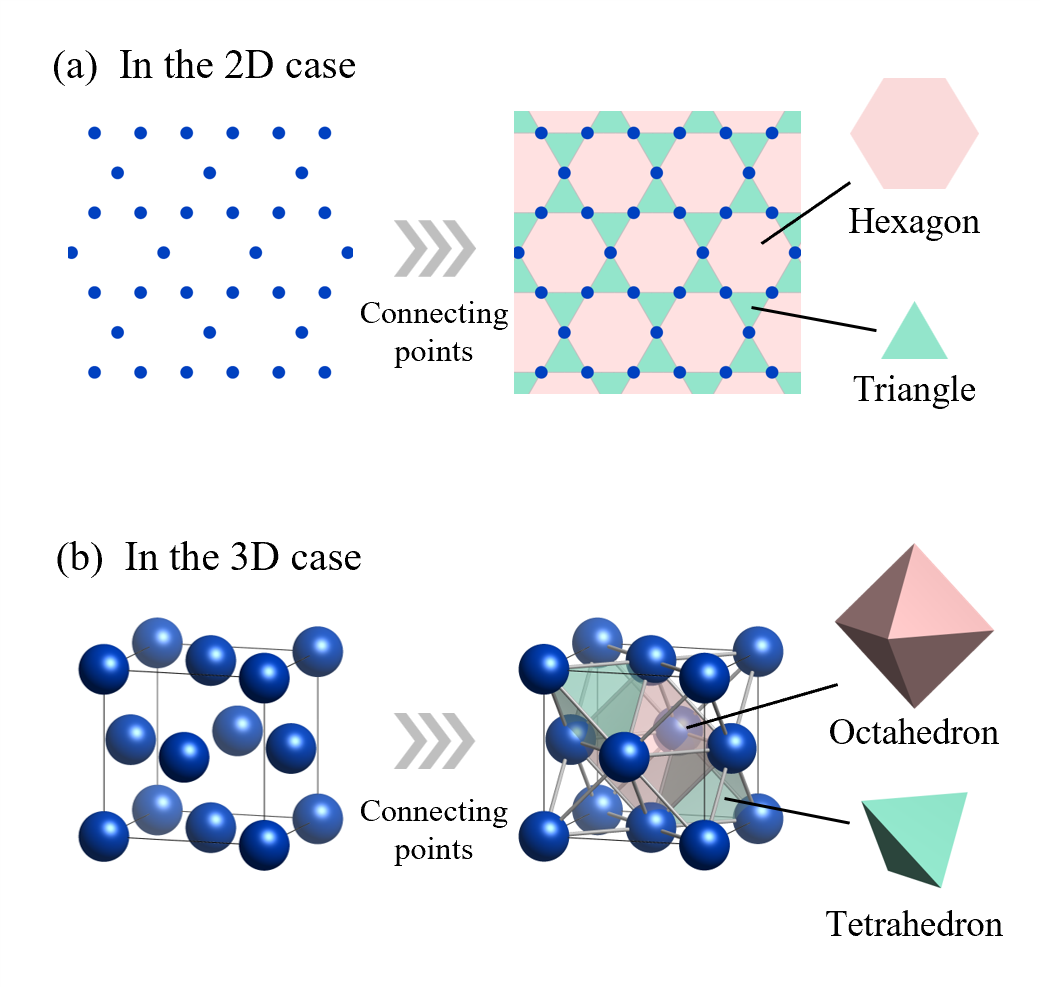}
  \caption{Examples of crystal structures represented as periodic arrangements of polyhedra. (a) A 2D kagome lattice. The blue dots represent lattice points (atomic sites), which form a tiling of triangles (green) and hexagons (pink) connected by edges. (b) A 3D face-centered cubic (FCC) lattice. The blue spheres represent atoms. The structure can be viewed as a space-filling tessellation of tetrahedra (green) and octahedra (pink).}
  \label{fig:ExCrystStruct}
\end{figure}

To understand relationships among crystal structures, it is essential to define their fundamental units. Just as prime numbers serve as the irreducible units of integers—since any integer can be expressed as a product of primes (e.g., $12 = 2 \times 2 \times 3$)—the basic units of composition are chemical elements. For instance, Li$_3$YCl$_6$ is composed of Li, Y, and Cl, and tuning their ratios or substituting elements forms the foundation of composition design.

So, what is the structural counterpart to an element in a crystal structure?
We argue that it is the space-filling polyhedron.
A crystal is essentially a periodic arrangement of unit cells, and each unit cell is composed of atoms arranged at the vertices of polyhedra that tile space without gaps.
For instance, a two-dimensional periodic lattice can be connected by edges to form a graph—a kagome lattice—consisting of triangles and hexagons that tile the plane as shown in Figure~\ref{fig:ExCrystStruct}a. These polygons serve as the minimal building blocks in 2D. 
Similarly, in 3D space, FCC structures can be regarded as tilings of space with tetrahedra and octahedra, where the vertices represent atomic positions as shown in Figure~\ref{fig:ExCrystStruct}b. Thus, just as elements serve as the minimal units of composition, polyhedra can serve as the minimal units of crystal structures.

The shape and connectivity of these polyhedra critically influence properties such as ionic conductivity. Theoretical studies indicate that the uniform tetrahedral tiling in BCC anion frameworks offers lower ion migration barriers compared to the mixed polyhedral networks found in FCC or HCP structures.\cite{Wang2015,Yokoyama2024} Indeed, the superionic conductor Li$_{10}$GeP$_2$S$_{12}$ features such a BCC-type framework.\cite{Kato2016}

\begin{figure}[htbp]
\centering
  \includegraphics[width=9cm]{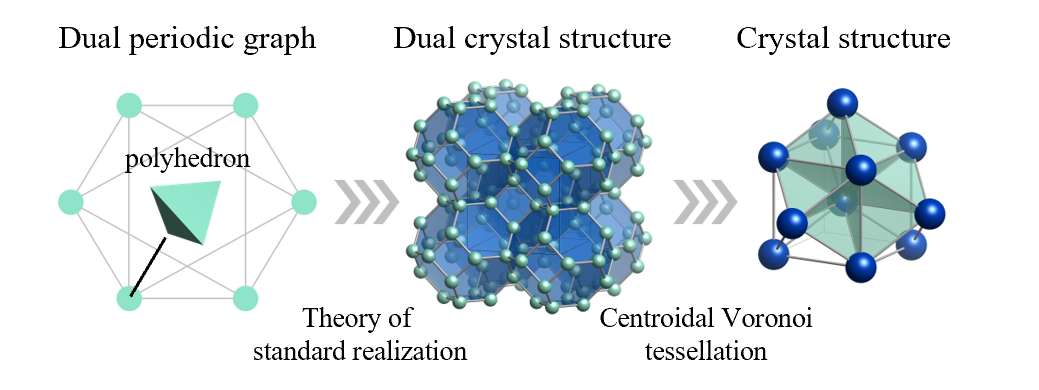}
  \caption{Conceptual diagram of the proposed crystal structure generation method. The process consists of three stages: (Left) The Dual Periodic Graph explicitly encodes the connectivity of the target polyhedra. Vertices represent polyhedral centers. (Middle) The Dual Crystal Structure is generated from the graph using the theory of standard realization. (Right) The final Crystal Structure is obtained by applying Centroidal Voronoi Tessellation (CVT) to the dual structure, placing atoms (blue spheres) at the vertices of the tiling. Reprinted with permission from ref.\cite{Yokoyama2023}  . Copyright 2024 American Chemical Society.}
  \label{fig:concept}
\end{figure}

Despite this importance, designing crystals by specifying constituent polyhedra—akin to building with Lego blocks—has been historically difficult. Conventional Crystal Structure Prediction (CSP) methods typically rely on stochastic searches of atomic coordinates and fail to explicitly encode polyhedral connectivity rules. To address this, we focus on a topological approach. In our recent work, we established a methodology integrating the theory of ``standard realization''\cite{Kotani2001}—a mathematical framework for determining the ideal, most symmetric embedding of a graph—with the concept of ``dual periodic graphs''\cite{Yokoyama2023} (Figure~\ref{fig:concept}). In a dual periodic graph, vertices represent polyhedral centers, and edges define their connectivity. By applying standard realization, we can deterministically generate high-symmetry crystal structures directly from topological input.

In this Highlight, we first review existing approaches to crystal structure prediction to contextualize our method within the broader landscape of computational materials science. 
We then detail the theoretical foundations of standard realization and dual periodic graphs. 
Subsequently, we demonstrate the effectiveness of this framework by reconstructing representative structures such as FCC, HCP, and BCC lattices. 
Finally, we discuss current challenges, such as handling complex systems, and future prospects for this topology-driven approach in materials discovery.

\section{Approaches to Crystal Structure Prediction}
The discovery of new crystal structures has traditionally relied on exploring the vast configuration space of atomic arrangements. 
Current computational strategies can be broadly categorized into three paradigms: global optimization approaches, deep generative models, and graph-theoretic approaches.

Conventionally, CSP is formulated as a global optimization problem on the potential energy surface. 
Stochastic search methods, including simulated annealing,\cite{pannetier1990prediction} evolutionary algorithms (e.g., USPEX\cite{Oganov2006uspex,Glass2006}), particle swarm optimization (e.g., CALYPSO\cite{Wang2012}), Bayesian optimization (e.g., CrySPY\cite{Yamashita2021}), and other methods \cite{wales1997global,martovnak2003predicting,amsler2010crystal,kusaba2022crystal,Chang2024,koshoji2024mathematical} have achieved remarkable success in identifying ground-state structures and stable high-pressure phases. 
However, these approaches typically rely on the random sampling of atomic coordinates.\cite{pickard2011ab,fredericks2021pyxtal}
Because they do not explicitly encode polyhedral connectivity rules a priori, they often expend significant computational resources exploring chemically unintuitive configurations where coordination environments are distorted or energetically unfavorable.
Even in approaches where clusters are randomly arranged to maintain local motifs, generating structures with desired polyhedra remains difficult.\cite{Yokoyama2021}

More recently, deep generative models have emerged as a new paradigm, enabling the direct proposal of crystal structures by learning statistical distributions from large crystallographic databases. 
Prominent examples include generative adversarial networks (e.g., CrystalGAN\cite{nouira2018crystalgan}), variational auto-encoders (e.g., iMatGen\cite{kim2019inverse}), diffusion models (e.g., CDVAE\cite{xie2021crystal}, DiffCSP\cite{jiao2023crystal}, MatterGen\cite{zeni2025generative}), and flow-based models (e.g., FlowMM\cite{miller2024flowmm}).
While these data-driven methods show great promise for accelerating materials discovery, their performance depends heavily on the quality and quantity of training data.
Furthermore, as probabilistic generators, they do not always guarantee strict geometric fidelity—such as high symmetry or precise polyhedral connectivity—without sophisticated constraints or adapter modules. 
Consequently, applying them to the deterministic ``inverse design'' of a specific target topology remains challenging.

To address these limitations, we focus on a distinct ``third pathway'': a deterministic approach based on topological crystallography.
A decisive advantage of this approach over the stochastic and data-driven methods described above is that it enables the ``exhaustive enumeration'' of structures within a mathematically defined search space.
By abstracting crystal structures as infinite periodic graphs (``nets''), it becomes possible to systematically explore the configuration space without bias towards known motifs.

However, a critical challenge in actually generating crystal structures via this approach is the ``decoding'' process: converting an abstract net into a crystal structure.
Initially, a topologically generated net is merely a graph representing vertex connectivity, lacking physical spatial coordinates.
To obtain a crystal structure, this abstract graph must be embedded into 3D Euclidean space.

This was rigorously achieved by the theory of ``standard realization,'' proposed by Kotani and Sunada in 2001\cite{Kotani2001}.
This method utilizes graph theory and variational principles to derive crystal structures from periodic graphs by minimizing an energy functional.
Unlike stochastic or data-driven methods, this framework allows for the direct construction of crystal structures that mathematically satisfy a target polyhedral topology.
Importantly, this framework has already been proven capable of solving non-trivial structural problems: it was used to mathematically reconstruct the diamond structure and to predict the K4 crystal.\cite{Itoh2009,Sunada2008}
Furthermore, Tagami \textit{et al.} applied this theory to generate Mackay-Terrones-like crystals (negatively curved cubic carbon crystals) with octahedral symmetry, demonstrating its ability to handle complex non-Bravais lattices defined by large unit cells.\cite{Tagami2014}

A similar approach, known as ``equilibrium placement'' (or barycentric embedding), was proposed by Delgado-Friedrichs and O'Keeffe in 2003.\cite{delgado2003identification}
This method determines a canonical structure by minimizing the harmonic energy of the edges, a principle that is mathematically homologous to standard realization.\cite{Sunada2012}
Leveraging this method, Foster \textit{et al.} exhaustively generated all possible structures under specified topological constraints and obtained their crystal structures via equilibrium placement.\cite{foster2004chemical}
This approach has successfully identified thousands of unique 4-connected nets, many of which correspond to known zeolite frameworks.

These topological methods have been instrumental in classifying crystal structures and establishing comprehensive databases.
Notable examples include the RCSR (Reticular Chemistry Structure Resource),\cite{okeeffe2008reticular} which curates fundamentally important topologies; EPINETS,\cite{ramsden2009three} which covers tiling patterns in Euclidean space; and the TTD (Topos Topological Database),\cite{blatov2009topological} a collection integrated into ToposPro\cite{blatov2014applied} that contains information on topological types of simple periodic nets and finite graphs.
Thus, the application of graph-theoretic approaches has flourished primarily in carbon materials and open-framework materials such as zeolites and Metal-Organic Frameworks (MOFs).
While there are interesting studies applying graph theory to phase transitions in binary metal crystals to predict high-pressure phases,\cite{kabanova2024topological} the application to dense metallic and ionic crystals remains limited.
For the design of such dense materials, the concept of space-filling polyhedra is essential, yet such an extension has not been fully realized.

To bridge this gap, we propose a new framework that incorporates space-filling polyhedra into the standard realization theory.
This allows for the rigorous generation of dense, highly symmetric crystal structures—such as FCC, HCP, and BCC—directly from polyhedral topology, offering a new pathway for the design of functional ionic and metallic materials.

\section{Standard Realizations}
\begin{figure}[h]
\centering
  \includegraphics[width=9cm]{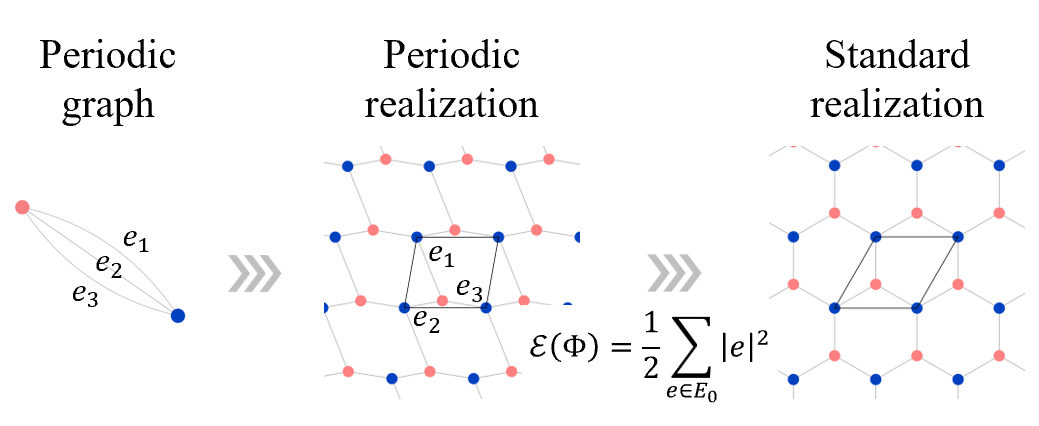}
  \caption{Illustration of the theory of standard realization. (Left) Periodic Graph: A finite graph consisting of two vertices (colored pink and blue for distinction) and three directed edges ($e_1$, $e_2$, $e_3$) that define the connectivity. (Middle) Periodic Realization: A geometric embedding of the graph into Euclidean space. The parallelogram (blue outline) represents the unit cell formed by the edge vectors. While this configuration satisfies the periodicity, it may not possess the highest possible symmetry. (Right) Standard Realization: The optimized configuration obtained by minimizing the harmonic energy function under a fixed-volume constraint. This process automatically yields the most symmetric unit cell shape and vertex positions (a hexagonal lattice in this 2D example).}
  \label{fig:StandReal}
\end{figure}

The theory of ``standard realization'' for crystal lattices, developed by Kotani and Sunada, formulates the crystal structure as a periodic realization of a finite graph in Euclidean space.\cite{Kotani2001}
The theory is based on graph theory and the principle of energy minimization as shown in Figure~\ref{fig:StandReal}.
It models atomic positions as vertices connected by harmonic springs (a harmonic oscillator model), and seeks the configuration that minimizes the total elastic energy under the constraint of fixed unit cell volume. 
This optimized configuration is termed the ``standard realization.''
In the following, the method to construct a standard realization is briefly described. 
More detailed mathematical descriptions can be found elsewhere.\cite{Kotani2001, Naito2009, Sunada2013, Naito2023}

In this framework, a crystal lattice is represented by a periodic extension (covering) of a finite periodic graph $X_0 = (V_0, E_0)$, where $V_0$ and $E_0$ denote the sets of vertices and edges, respectively. A spanning tree $X_T = (V_0, E_T)$ is then extracted from $X_0$, which is a subgraph that contains all vertices but no closed paths. The first Betti number $b$ of the graph, representing the number of independent closed paths, is defined as $b = |E_0| - |E_T|$.

Each edge in $E_0$ is assigned a direction, and $b$ linearly independent closed paths $\alpha_1, \, \alpha_2, \, \ldots, \, \alpha_b$ are chosen to span the first homology group $H_1(X_0, \mathbb{Z})$. The $b$-dimensional real vector space $H_1(X_0, \mathbb{R})$ consists of linear combinations of these closed paths with real coefficients.

We then define an inner product $\langle e_j, e_k \rangle = \delta_{jk}$ for each pair of edges and compute two matrices: the $N \times b$ matrix $B = [\langle e_j, \alpha_k \rangle]$, and the $b \times b$ Gram matrix $G_0 = [\langle \alpha_j, \alpha_k \rangle]$. From these, we obtain the coefficient matrix $A = G_0^{-1} B$, which expresses each edge vector in terms of the chosen closed-path basis.

When the Betti number $b$ exceeds the spatial dimension $d$, a projection onto a suitable $d$-dimensional subspace is required. This choice is made by selecting $d$ out of the $b$ closed paths, typically those forming a subspace that captures the symmetry of interest.

The lattice vectors $p_x, p_y, p_z$ (in 3D) are constructed from the projection of $G_0$, ensuring that the lengths and angles between basis vectors match the standard realization conditions. Once the edge vectors are determined, we compute the position of each vertex by tracing paths through the spanning tree, starting from a reference vertex placed at the origin. This yields the atomic positions for the unit cell.

Finally, the unit cell is extended periodically using the lattice vectors to generate the infinite crystal structure. Among all possible periodic realizations of the graph, the standard realization has the highest spatial symmetry, faithfully reflecting the inherent symmetry of the underlying graph.

\begin{figure}[h]
\centering
  \includegraphics[width=9cm]{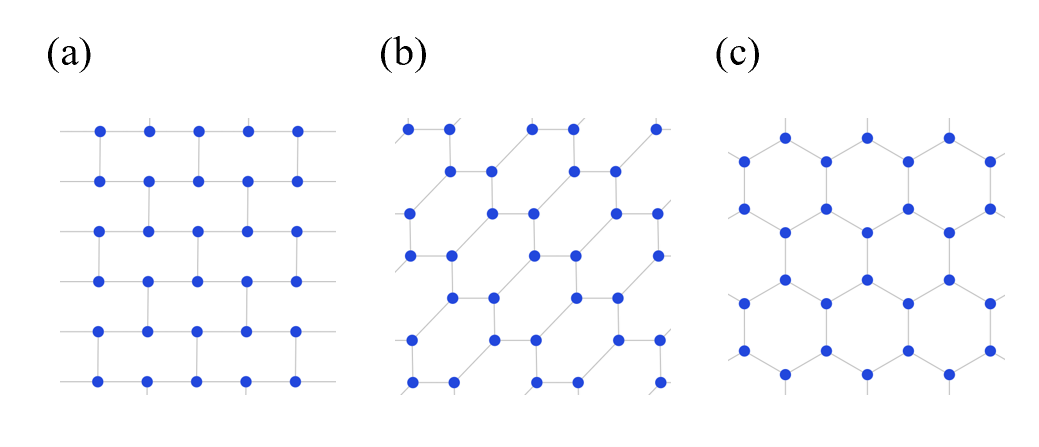}
  \caption{Three different realizations of the same periodic hexagonal graph. (a, b) Examples of periodic realizations that satisfy the topological connectivity but exhibit lower symmetry. (c) The standard realization, which uniquely yields the configuration with the highest symmetry.}
  \label{fig:ExReal}
\end{figure}

An illustrative example is shown in Figure~\ref{fig:ExReal}, where different realizations of a hexagonal lattice are compared. The most symmetric case corresponds to the hexagonal tiling, which is obtained as the standard realization.

\section{Dual Crystal Structure and Dual Periodic Graph}
In general, a dual relationship refers to a correspondence between two paired entities—each being the dual of the other. For instance, an octahedron and a cube, or a Delaunay diagram and a Voronoi diagram, are well-known dual pairs. The dual of a dual structure returns to the original structure.

\begin{figure}[h]
\centering
  \includegraphics[width=9cm]{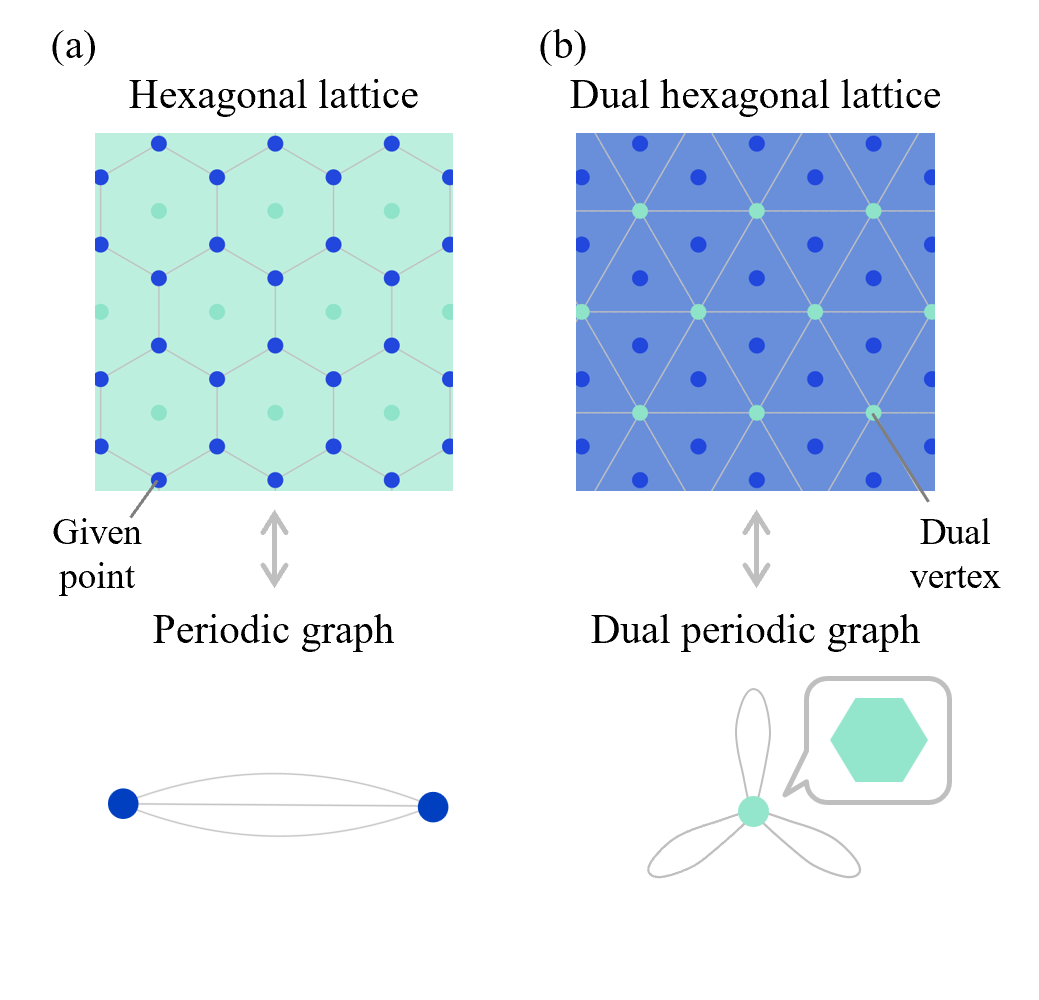}
  \caption{Relationship between the hexagonal lattice and its dual representation in two dimensions. (a) The hexagonal lattice (top) and its periodic graph (bottom). (b) The dual crystal structure (top) and the corresponding dual periodic graph (bottom). In the dual periodic graph, the dual vertex (center of a hexagon) is connected to six edges, reflecting the six sides of the original hexagonal tile.}
  \label{fig:Ex2dLattice}
\end{figure}

Figure~\ref{fig:Ex2dLattice}a shows an example of a two-dimensional (2D) crystal structure and its corresponding dual crystal structure. The structure illustrated is the hexagonal lattice. Here, a $b$-dimensional crystal structure is defined as a set of points that are periodically repeated in $b$-dimensional space. When edges are added between specific points, this 2D structure forms a tessellation consisting of a hexagon. However, this polygonal composition is not apparent from the original periodic graph alone, which merely captures the connectivity of atomic sites.

To explicitly represent these tiles in a graph-theoretic framework, we introduce the concept of a dual crystal structure. A dual crystal structure is formed by connecting the centers of the space-filling polyhedra that constitute the original crystal structure. For instance, the dual of the hexagonal lattice is constructed by linking the centers of the hexagons that tile the plane.

The dual periodic graph is then defined as the periodic graph derived from this dual crystal structure. In the case of the hexagonal lattice, the dual vertex in Figure~\ref{fig:Ex2dLattice}a has six connecting edges—corresponding to the number of edges in a hexagon.

Importantly, the original crystal structure and its dual share the same spatial symmetry when derived from a consistent polyhedral tiling.
Therefore, the quotient group of periodicity, $X/X_0$, remains invariant between the original and dual graphs.

\begin{figure}[h]
\centering
  \includegraphics[width=9cm]{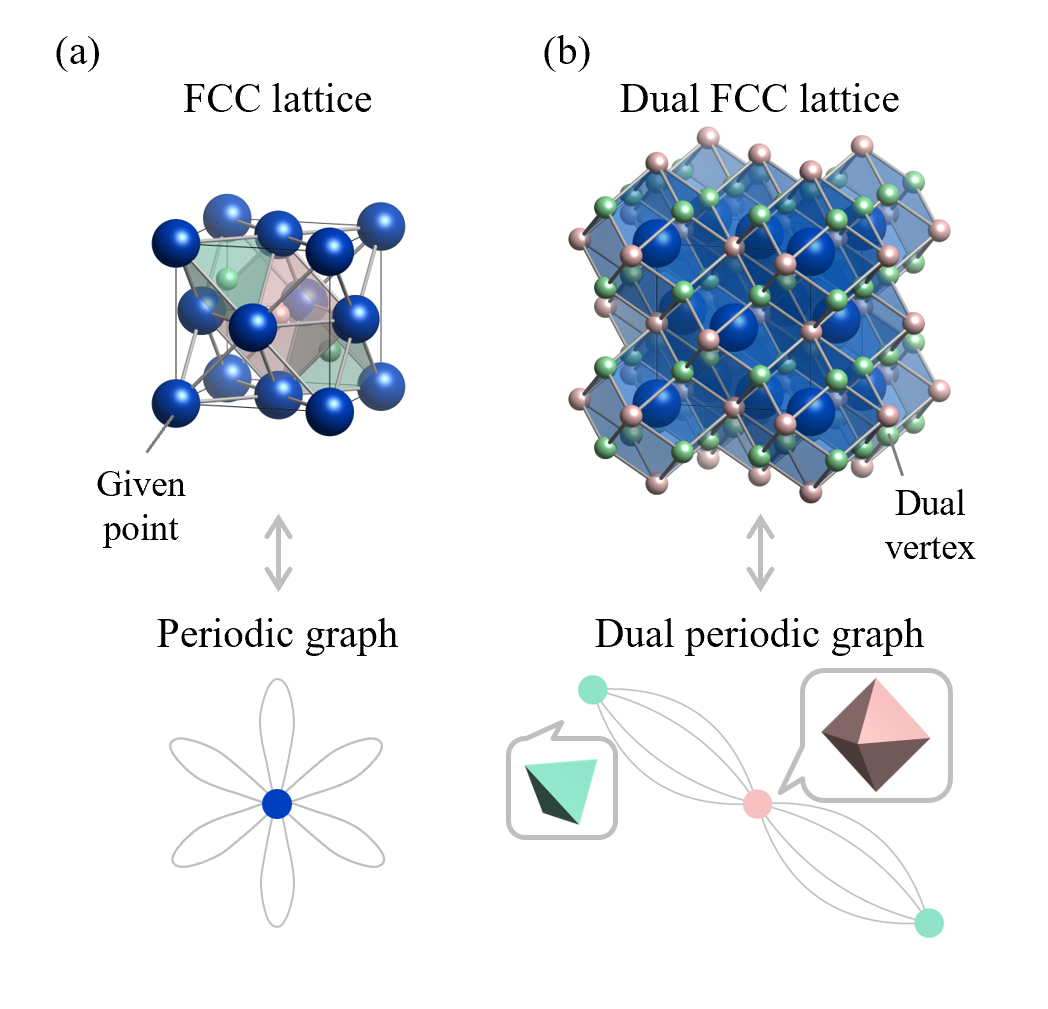}
  \caption{Relationship between the Face-Centered Cubic (FCC) lattice and its dual representation. (a) FCC lattice (top) and its periodic graph (bottom).  (b) The dual crystal structure (top) and the corresponding dual periodic graph (bottom). In the dual periodic graph, the vertices represent the centers of tetrahedra and octahedra: the red vertex corresponds to an octahedral site (connected to 8 edges), and green vertices correspond to tetrahedral sites (each connected to 4 edges).  Reprinted with permission from ref.\cite{Yokoyama2023}. Copyright 2024 American Chemical Society.}
  \label{fig:Ex3dLattice}
\end{figure}

Just as in the 2D case, dual crystal structures and dual periodic graphs can also be defined in three-dimensional (3D) space. 
Figure~\ref{fig:Ex3dLattice} shows the FCC structure and its dual. 
By connecting adjacent atomic sites, the primitive cell of this 3D structure is seen to consist of two tetrahedra and one octahedron.
The dual crystal structure is obtained by connecting the centers of these polyhedra. 
Although the periodic graph of the FCC structure does not reveal this polyhedral information, its dual periodic graph does reveal this: the green vertex connects to four edges and the red vertex to eight, corresponding to the number of faces in a tetrahedron and octahedron, respectively.

Thus, a dual periodic graph encodes the space-filling polyhedral framework of a crystal structure. 
By applying the standard realization method to such graphs, one can reconstruct a dual crystal structure that inherently reflects the intended tiling geometry.

Although a dual graph is not uniquely determined for an abstract graph alone, it becomes uniquely defined when the spatial positions of graph vertices are specified. 
Here, “uniquely defined” means that the dual graph is determined uniquely up to the symmetry operations of the polyhedral tiling.
When the spatial embedding is fixed the adjacency relations among polyhedral cells become unambiguous, yielding a canonical dual representation.
Our method begins by defining atomic bonds in the crystal structure, then forms a dual crystal structure based on these bonds and vertices. From this, the dual periodic graph is extracted. 
In other words, the dual periodic graph is defined by the geometry of a polyhedral tiling derived from the original structure.

To generate such tilings, we employ centroidal Voronoi tessellation (CVT), which partitions space by assigning regions to atoms based on proximity. 
In 2D, this yields polygons; in 3D, polyhedra. 
By treating the intersections of these regions as dual vertices, CVT enables bidirectional conversion between crystal and dual crystal structures.
CVT ensures that the resulting polyhedral cells are centroidal and space-filling, which provides a geometrically consistent bridge between the dual embedding and the atomic positions in the final structure.

\section{Structure Generation Framework}
\begin{figure}[h]
\centering
  \includegraphics[width=9cm]{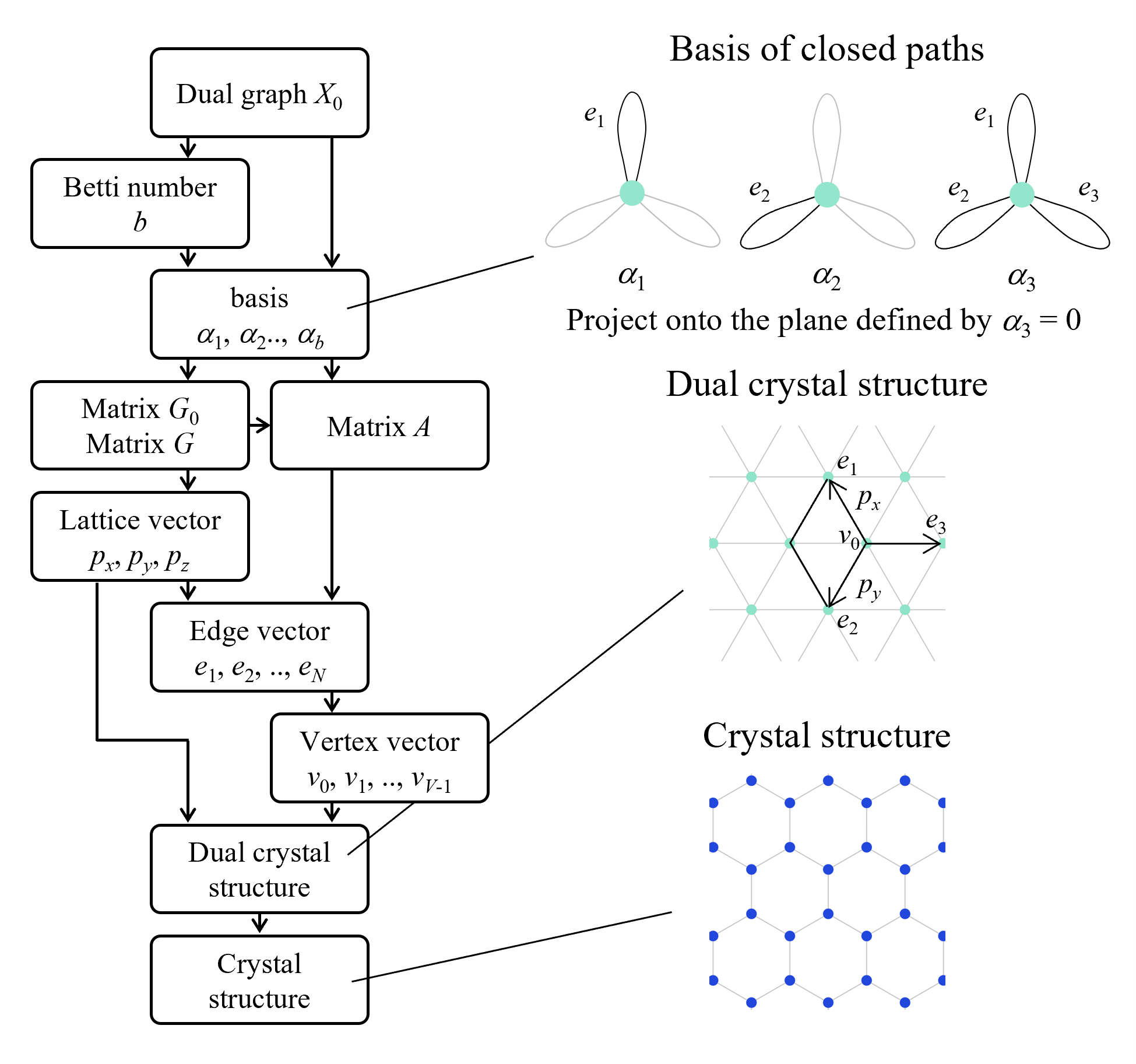}
  \caption{Flowchart of the crystal structure generation method and the step-by-step generation of a hexagonal lattice from its dual periodic graph. Reprinted with permission from ref.\cite{Yokoyama2023}. Copyright 2024 American Chemical Society.}
  \label{fig:FlowChart}
\end{figure}

In this section, we explain the detailed procedure for generating a crystal structure from a given dual periodic graph (see Figure~\ref{fig:FlowChart}).

The input to the method is a dual periodic graph that encodes the shape and connectivity of space-filling polyhedra. The corresponding crystal structure is generated in eight steps. Steps 1 through 7 involve constructing a symmetric dual structure based on the theory of standard realization. Step 8 then converts this dual structure into a crystal structure via geometric transformation. These steps
are conducted as follows.

\begin{enumerate}
\item Calculate the Betti number $b$ from the given dual periodic graph $X_0$. Here, $b$ is equal to the number of edges in the graph $|E_0|$ minus the number of edges of maximal spanning trees $|E_T|$.
\item Define a basis set $\{\alpha_j\}_{j=1}^b$ based on the number of selected closed paths corresponding to the Betti number $b$ obtained in Step 1. When $b > 3$, the choice of basis determines the three-dimensional subspace $\{\alpha_j\}_{j=1}^3$ onto which the structure will be projected, and the projection is taken onto the subspace defined by setting $\alpha_4 = \cdots = \alpha_b = 0$.
\item Calculate the matrices $G_0$, $G$, and $A$ from the basis set $\{\alpha_j\}_{j=1}^b$ obtained in step 2 to project from the $b$-dimensional vector space to the $3$D vector subspace. The matrices $G_0$ and $G$ represent the conditions satisfied by the $b$- and $3$D lattice vectors, respectively. The matrix $A$ represents the basis of the edges in the $b$-dimensional vector space.
\item Define lattice vectors $p_x$, $p_y$, and $p_z$ such that the matrix $G$ obtained in step 3 is satisfied. The lattice vectors can also be obtained by Cholesky decomposition of the matrix $G$.
\item Calculate the edge vectors $\{e_j\}_{j=1}^N$ from the matrix $A$ obtained in step 3 and the lattice vectors $p_x$, $p_y$, and $p_z$ obtained in step 4.
\item Calculate the vertex vectors $\{v_k\}_{k=0}^{|V|-1}$ from the lattice vectors $p_x$, $p_y$, and $p_z$ obtained in step 4 and the edge vectors $\{e_j\}_{j=1}^N$ obtained in step 5. The vertex vectors correspond to the primitive coordinates.
\item Generate a dual crystal structure from the lattice vectors $p_x$, $p_y$, $p_z$ obtained in step 4 and the vertex vectors $\{v_k\}_{k=0}^{|V|-1}$ obtained in step 6.
\item Transform the dual structure obtained in step 7 into a crystal structure by CVT.
\end{enumerate}

We begin by applying this process to a hexagonal lattice in two dimensions. 
This example illustrates the full procedure of standard realization. 
Subsequently, we apply the method to the dual periodic graph of the hexagonal lattice and demonstrate how the hexagonal lattice can be reconstructed from its dual representation.

\subsection{Case Study: Hexagonal Lattice}
This subsection presents the standard realization of a hexagonal lattice in two dimensions based on the original periodic graph—rather than the dual periodic graph—following a seven-step procedure that omits Step 8.

Step 1: We begin by computing the Betti number $b$, which represents the number of independent closed paths in the graph. 
It is given by the total number of edges $|E_0|$ minus the number of edges in a spanning tree $|E_T|$. 
For the original graph of the hexagonal lattice (Figure~\ref{fig:Ex2dLattice}a), we have $|E_0| = 3$ and $|E_T| = 1$, resulting in $b = 2$. Since the structure is to be realized in two-dimensional space, the spatial dimension is $d = 2$.

Step 2: Next, we define the basis set based on the number of selected closed paths corresponding to $b = 2$. 
Let $e_1$, $e_2$, and $e_3$ denote the three directed edges of the graph, oriented from the left to the right vertex. 
Based on these, we define the basis set as

\begin{align*}
\alpha_1 = e_1 -e_3, \quad \alpha_2 = e_2 - e_3.
\end{align*}

Step 3: Using this basis set, we compute the inner product matrix $G_0$ and the matrix $B$.
Since the Betti number $b$ equals the spatial dimension $d$ in this case, we set $G = G_0$.
The matrix $G$ encodes the target geometry, specifically representing the inner products among the basis elements of the first homology group $H_1(X_0, \mathbb{Z})$.
Accordingly, $G$ determines the lengths and angles of the lattice basis vectors in the projected space.
The matrix $G$ is computed as follows:

\begin{equation}
  G=  
  G_0=
  \begin{bmatrix}
    |\alpha_1|^2 & \inner{\alpha_1}{\alpha_2} \\
    \inner{\alpha_2}{\alpha_1} & |\alpha_2|^2 \\
  \end{bmatrix}
  =
  \begin{bmatrix}
    2 & 1  \\
    1 & 2  \\
  \end{bmatrix}.
\end{equation}

The matrix $B$ is constructed from the inner products between the edge vectors and the basis set as follows:
\begin{equation}
  B=  
  \begin{bmatrix}
    \inner{e_1}{\alpha_1} & \inner{e_1}{\alpha_2}\\
    \inner{e_2}{\alpha_1} & \inner{e_2}{\alpha_2}\\
    \inner{e_3}{\alpha_1} & \inner{e_3}{\alpha_2}\\
  \end{bmatrix}
  =
  \begin{bmatrix}
    1 & 0  \\
    0 & 1  \\
    -1 & -1  \\
  \end{bmatrix}.
\end{equation}
The matrix $A = G^{-1}B$ expresses each edge vector in terms of the basis. 
By performing the matrix multiplication, we obtain:
\begin{align*}
e_1 = \frac{2}{3}\alpha_1 -\frac{1}{3}\alpha_2, \quad e_2 = -\frac{1}{3}\alpha_1 +\frac{2}{3}\alpha_2, \quad e_3 = -\frac{1}{3}\alpha_1 -\frac{1}{3}\alpha_2.
\end{align*}

Step 4: we determine the lattice basis vectors $p_x$ and $p_y$ in 2D space so that they satisfy the geometric conditions encoded in $G$. 
Specifically, the following relationship must hold:
\begin{equation}
  G=  
  \begin{bmatrix}
    |p_x|^2 & \inner{p_x}{p_y} \\
    \inner{p_y}{p_x} & |p_y|^2 \\
  \end{bmatrix}.
  \label{eq::G}
\end{equation}
From this condition and the form of $G$ obtained in Step 3, we observe that the standard realization of the hexagonal lattice yields translation vectors of equal length enclosing an angle of $60^\circ$.
One possible set of lattice vectors satisfying this condition is:
\begin{align*}
p_x = 
    \begin{bmatrix}
    \sqrt2\\
    0\\
    \end{bmatrix}, \quad
p_y = 
    \begin{bmatrix}
    -1/\sqrt2\\
    -\sqrt{3/2}\\
    \end{bmatrix}.
\end{align*}
Note that the choice of lattice vectors is not unique due to rotational and reflectional symmetry; they can also be computed using numerical methods such as Cholesky decomposition.

Step 5: Using the coefficient matrix $A$ and the chosen lattice basis vectors, we can express the edge vectors explicitly as

\begin{align*}
e_1 = 
    \begin{bmatrix}
    1/\sqrt2\\
    1/\sqrt6\\
    \end{bmatrix}, \quad
e_2 = 
    \begin{bmatrix}
    0\\
    \sqrt{2/3}\\
    \end{bmatrix}, \quad
e_3 = 
    \begin{bmatrix}
    -1/\sqrt{2}\\
    -1/\sqrt{6}\\
    \end{bmatrix}.
\end{align*}

Step 6: We place the reference vertex $v_0$ at the origin and define the other vertex positions as $v_0 + e_1$, $v_0 + e_2$, $v_0 + e_3$. 

Step 7: We generate the crystal structure by periodically replicating the unit triangle defined by $v_0$, $v_1$, and $v_2$ using the lattice vectors $p_x$ and $p_y$. 
Replicating this triangle periodically using $p_x$ and $p_y$ yields the hexagonal lattice shown in Figure~\ref{fig:Ex2dLattice}a. 

While the construction is mathematically complete, the equivalence between this structure and a hexagonal tiling becomes visually evident only when the diagram is drawn.

\subsection{Case Study: The Dual of Hexagonal Lattice}
To illustrate how polyhedral shapes can be explicitly considered, we next apply the method to the dual periodic graph of the hexagonal lattice (Figure~\ref{fig:Ex2dLattice}b). 
In this dual periodic graph, each vertex is connected to six edges, reflecting the six edges of a hexagon.

Step 1: We compute the Betti number $b$. This graph has $|E_0| = 3$ and $|E_T| = 0$, giving a Betti number of $b = 3$. Since the realization is in 2D space, the target spatial dimension is $d = 2$.

Step 2: Let $e_1$, $e_2$, and $e_3$ be directed edges forming three closed paths. 
If we let $\alpha_1 = e_1$, $\alpha_2 = e_2$, and $\alpha_3 = e_3$, the resulting 3D standard realization corresponds to a cubic lattice. However, here we take:
\begin{align*}
\alpha_1 = e_1, \quad \alpha_2 = e_2, \quad \alpha_3 = e_1 + e_2 + e_3,
\end{align*}
and construct the standard realization in the 2D subspace spanned by $\alpha_1$ and $\alpha_2$. Geometrically, this corresponds to projecting the cubic lattice onto its (111) plane.

Step 3: We compute the inner product matrix $G_0$, the edge-closed path matrix $B$, and the coefficient matrix $A$ as follows:
\begin{align*}
G_0 = \begin{bmatrix} 
        1 & 0 & 1 \\
        0 & 1 & 1 \\
        1 & 1 & 3 \\
    \end{bmatrix}, \quad
B = \begin{bmatrix} 
        1 & 0 & 1 \\
        0 & 1 & 1 \\
        0 & 0 & 1 \\
    \end{bmatrix}, \quad
A = G_0^{-1}B=\begin{bmatrix} 
        1 & 0 & 0 \\
        0 & 1 & 0 \\
        -1 & -1 & 1 \\
    \end{bmatrix},
\end{align*}
The three edge vectors in 3D become:
\begin{align*}
e_1 = \alpha_1, \quad e_2 = \alpha_2, \quad e_3 = -\alpha_1 - \alpha_2 + \alpha_3.
\end{align*}
Since $G_0$ describes the inner products of the 3D basis vectors before projection, we must recalculate the projected metric matrix $G$ for the (111) plane. 
If we partition $G_0$ as:
\begin{equation}
G_0 = \begin{bmatrix} G_{11} & G_{12} \\ G_{21} & G_{22} 
\label{eq::G0}
\end{bmatrix},
\end{equation}
then the effective 2D metric is:
\begin{equation}
G = G_{11} - G_{12} G_{22}^{-1} G_{21}.
\label{eq::G_G0}
\end{equation}
Here, $G_{11}$ is a $d \times d$ matrix, $G_{12}$ is $d \times (b-d)$, $G_{21}$ is $(b-d) \times d$, and $G_{22}$ is $(b-d) \times (b-d)$. In this case, $G$ is given by
\begin{align*}
G = \frac{1}{3} \begin{bmatrix}2&-1 \\ -1&2 \end{bmatrix},
\end{align*}

Step 4: Using the matrix $G$, we compute the projected lattice vectors $p_x$ and $p_y$ in accordance with Eq.~\ref{eq::G},
\begin{align*}
p_x = \frac{1}{2}
    \begin{bmatrix}
      1\\
      -\sqrt3\\
    \end{bmatrix}, \quad
p_y = \frac{1}{2}
    \begin{bmatrix}
      1\\
      \sqrt3\\
    \end{bmatrix}.
\end{align*}

Step 5: We reconstruct the edge vectors  $e_1$, $e_2$, $e_3$ as follows:
\begin{align*}
e_1=p_x = \frac{1}{2}
    \begin{bmatrix}
      1\\
      -\sqrt3\\
    \end{bmatrix}, \quad
e_2=p_y = \frac{1}{2}
    \begin{bmatrix}
      1\\
      \sqrt3\\
    \end{bmatrix}, \quad
e_3=-p_x-p_y = \frac{1}{2}
    \begin{bmatrix}
      1\\
      0\\
    \end{bmatrix}.
\end{align*}

Step 6 and 7: Using $p_x$, $p_y$, and a reference vertex $v_0$, we construct the dual crystal structure by periodically replicating the unit. 
The resulting structure is shown in Figure~\ref{fig:Ex2dLattice}b.

Step 8: Finally, we apply CVT to the dual crystal structure to obtain the crystal structure. 
By placing new vertices at the centers of the triangles and connecting them appropriately, the original hexagonal tiling is recovered.

By employing a dual periodic graph, crystal structures can be systematically generated from prescribed polyhedral information, as demonstrated by the reconstruction of the hexagonal lattice via the standard realization method.

\section{Applications to Representative Structures: FCC, HCP, and BCC}
\begin{figure*}[htbp]
\centering
  \includegraphics[width=\textwidth]{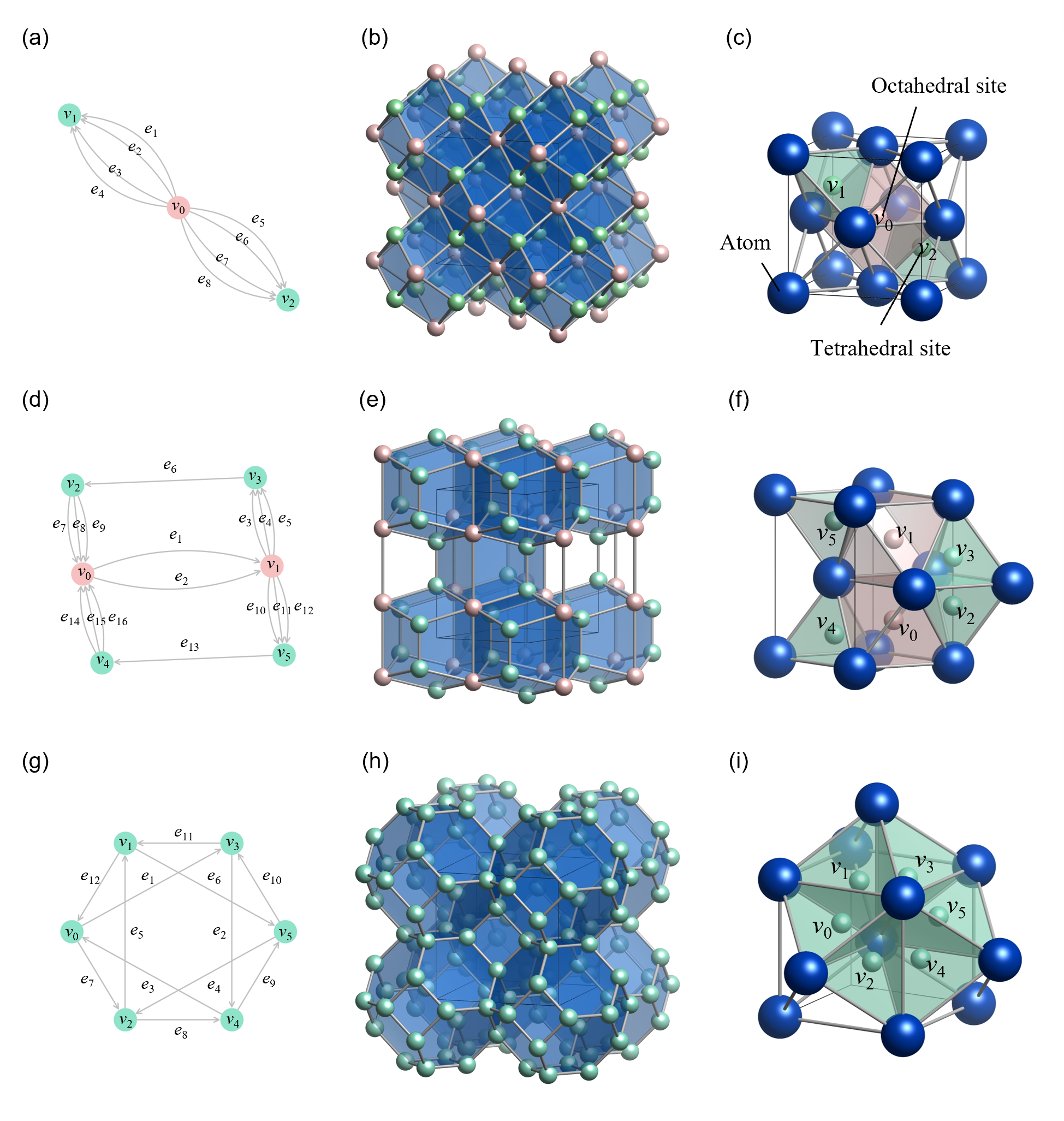}
  \caption{Reconstruction of representative crystal structures using the proposed method. The rows correspond to (a-c) Face-Centered Cubic (FCC), (d-f) Hexagonal Close-Packed (HCP), and (g-i) Body-Centered Cubic (BCC) structures. (Left Column: a, d, g) Dual periodic graphs. The vertices represent the centers of voids (interstitial sites) in the target crystal: red vertices correspond to octahedral sites, and green vertices correspond to tetrahedral sites. The edges denote the connectivity between these polyhedra. (Middle Column: b, e, h) Dual crystal structures generated by standard realization. These show the space-filling packing of polyhedra (e.g., rhombic dodecahedra for FCC, truncated octahedra for BCC) derived mathematically from the graphs. (Right Column: c, f, i) Final crystal structures obtained via CVT. Blue spheres indicate atomic positions, reconstructing the target FCC, HCP, and BCC lattices with high fidelity. Reprinted with permission from ref.\cite{Yokoyama2023}. Copyright 2024 American Chemical Society.}
  \label{fig:FccHcpBcc}
\end{figure*}

Following the same procedure used for two-dimensional lattices in the previous section, a three-dimensional crystal structure can be generated from a dual peridoc graph. 
Here, we describe the specific steps using the FCC structure as an example. 
The dual periodic graph of the FCC structure is shown in Figure~\ref{fig:FccHcpBcc}a.

The primitive cell of the FCC structure contains two interstitial tetrahedral sites and one interstitial octahedral site.
In other words, the FCC structure can be viewed as a tiling of two tetrahedra and one octahedron. 
Each tetrahedron shares all of its faces with octahedra, and each octahedron shares all of its faces with tetrahedra.
In the dual periodic graph, the green vertices ($v_1$ and $v_2$) are connected to four edges, and the red vertex ($v_0$) is connected to eight edges. 
These correspond to the number of faces of the tetrahedron and octahedron, respectively.

We now follow the procedure to generate the FCC structure from its dual periodic graph. 

Step 1: We compute the Betti number $b$. The dual periodic graph has 8 edges and a spanning tree with 2 edges, resulting in $b = |E_0| - |E_T| = 8 - 2 = 6$. Since the structure is embedded in three-dimensional space, the dimension is $d = 3$.

\begin{figure}[h]
\centering
  \includegraphics[width=9cm]{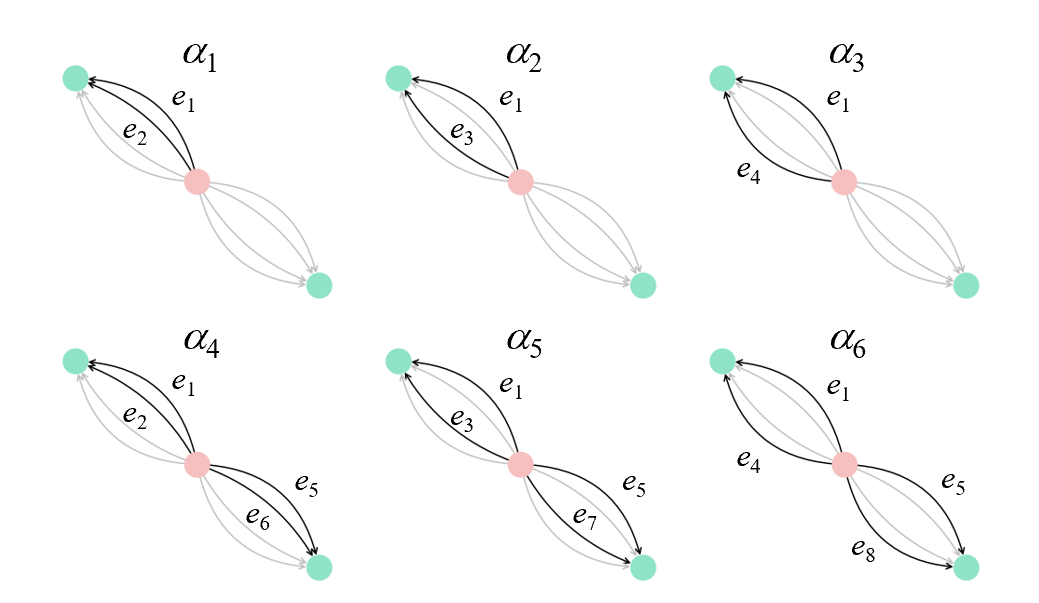}
  \caption{Basis of closed paths in the dual periodic graph of the face-centered cubic (FCC) structure used for standard realization. The graph contains six independent closed paths ($\alpha_1$ through $\alpha_6$), corresponding to the Betti number $b=6$. The vertices represent the centers of the space-filling polyhedra: the red vertex corresponds to an octahedral site, and green vertices correspond to tetrahedral sites. Reprinted with permission from ref.\cite{Yokoyama2023}. Copyright 2024 American Chemical Society.}
  \label{fig:FccClosedPath}
\end{figure}

Step 2: We define the basis set based on the number of
selected closed paths corresponding to $b = 6$.
Let the eight edges of the basic graph be denoted as $e_1$ through $e_8$. 
We define the closed path basis as follows:
\begin{align*}
\alpha_1 &= e_1 - e_2, \\
\alpha_2 &= e_1 - e_3, \\
\alpha_3 &= e_1 - e_4, \\
\alpha_4 &= e_1 - e_2 + e_5 - e_6, \\
\alpha_5 &= e_1 - e_3 + e_5 - e_7, \\
\alpha_6 &= e_1 - e_4 + e_5 - e_8.
\end{align*}
These correspond to the closed paths illustrated in Figure~\ref{fig:FccClosedPath}.

Step 3: Using this basis, we compute the matrices $G_0$, $B$, and $A$.
The matrix $G_0$ is given by:
\begin{align*}
G_0 = \begin{bmatrix} 
2 & 1 & 1 & 2 & 1 & 1\\
1 & 2 & 1 & 1 & 2 & 1\\
1 & 1 & 2 & 1 & 1 & 2\\
2 & 1 & 1 & 4 & 2 & 2\\
1 & 2 & 1 & 2 & 4 & 2\\
1 & 1 & 2 & 2 & 2 & 4\\
\end{bmatrix}.
\end{align*}
The matrix $B$ is:
\begin{align*}
B = \begin{bmatrix}
1 & 1 & 1 & 1 & 1 & 1\\
-1 & 0 & 0 & -1 & 0 & 0\\
0 & -1 & 0 & 0 & -1 & 0\\
0 & 0 & -1 & 0 & 0 & -1\\
0 & 0 & 0 & 1 & 1 & 0\\
0 & 0 & 0 & -1 & 0 & 0\\
0 & 0 & 0 & 0 & -1 & 0\\
0 & 0 & 0 & 0 & 0 & -1\\
\end{bmatrix}.
\end{align*}
Multiplying $G^{-1}B$ yields the matrix $A$:
\begin{align*}
A = \frac{1}{4}\begin{bmatrix} 
1 & 1 & 1 & 0 & 0 & 0\\
-3 & 1 & 1 & 0 & 0 & 0\\
1 & -3 & 1 & 0 & 0 & 0\\
1 & 1 & -3 & 0 & 0 & 0\\
-1 & -1 & -1 & 1 & 1 & 1\\
3 & -1 & -1 & -3 & 1 & 1\\
-1 & 3 & -1 & 1 & -3 & 1\\
-1 & -1 & 3 & 1 & 1 & -3\\
\end{bmatrix}.
\end{align*}
With Eqs.~\ref{eq::G0} and \ref{eq::G_G0}, the matrix $G$ in 3D space is computed as:
\begin{align*}
G = \frac{1}{2}\begin{bmatrix} 
2 & 1 & 1\\
1 & 2 & 1\\
1 & 1 & 2\\
\end{bmatrix}.
\end{align*}

Step 4: We determine lattice vectors $p_x$, $p_y$, and $p_z$ satisfying:
\begin{equation}
  G=  
  \begin{bmatrix}
    |p_x|^2 & \inner{p_x}{p_y} & \inner{p_x}{p_z} \\
    \inner{p_y}{p_x} & |p_y|^2 & \inner{p_y}{p_z} \\
    \inner{p_z}{p_x} & \inner{p_z}{p_y} & |p_z|^2 \\
  \end{bmatrix}.
\end{equation}
An example set of lattice vectors is:
\begin{align*}
p_x = \frac{1}{2}
    \begin{bmatrix}1\\0\\1\\\end{bmatrix}, \quad
p_y = \frac{1}{2}
    \begin{bmatrix}1\\1\\0\\\end{bmatrix}, \quad
p_z = \frac{1}{2}
    \begin{bmatrix}0\\1\\1\\\end{bmatrix}.
\end{align*}

Step 5: Using the matrix $A$ and the lattice vectors, we obtain the edge vectors $e_1$ through $e_8$ as:
\begin{alignat*}{3}
  e_1&=\frac{1}{4}p_x+\frac{1}{4}p_y+\frac{1}{4}p_z, &\quad
  e_2&=-\frac{3}{4}p_x+\frac{1}{4}p_y+\frac{1}{4}p_z, \\
  e_3&=\frac{1}{4}p_x-\frac{3}{4}p_y+\frac{1}{4}p_z, &\quad
  e_4&=-\frac{3}{4}p_x+\frac{1}{4}p_y-\frac{3}{4}p_z, \\
  e_5&=-\frac{1}{4}p_x-\frac{1}{4}p_y-\frac{1}{4}p_z, &\quad
  e_6&=\frac{3}{4}p_x-\frac{1}{4}p_y-\frac{1}{4}p_z, \\
  e_7&=-\frac{1}{4}p_x+\frac{3}{4}p_y-\frac{1}{4}p_z, &\quad
  e_8&=-\frac{1}{4}p_x-\frac{1}{4}p_y+\frac{3}{4}p_z, \\
\end{alignat*}
We then define vertex vectors from these edges. 
\begin{align*}
e_1 = \frac{1}{2}
    \begin{bmatrix}1\\1\\1\\\end{bmatrix}, \quad
e_2 = \frac{1}{2}
    \begin{bmatrix}-1\\1\\-1\\\end{bmatrix}, \quad
e_3 = \frac{1}{2}
    \begin{bmatrix}-1\\-1\\1\\\end{bmatrix}, \quad
e_4 = \frac{1}{2}
    \begin{bmatrix}1\\-1\\-1\\\end{bmatrix}, \\
e_5 = \frac{1}{2}
    \begin{bmatrix}-1\\-1\\-1\\\end{bmatrix}, \quad
e_6 = \frac{1}{2}
    \begin{bmatrix}1\\-1\\1\\\end{bmatrix}, \quad
e_7 = \frac{1}{2}
    \begin{bmatrix}1\\1\\-1\\\end{bmatrix}, \quad
e_8 = \frac{1}{2}
    \begin{bmatrix}-1\\1\\1\\\end{bmatrix}.
\end{align*}

Step 6: Taking $v_0$ as the origin, $v_1 = v_0 + e_1$, and $v_2 = v_0 + e_5$, the three vertex vectors are:

\begin{align*}
v_0 = \frac{1}{2}
    \begin{bmatrix}0\\0\\0\\\end{bmatrix}, \quad
v_1 = v_0 + e_1 = \frac{1}{2}
    \begin{bmatrix}1\\1\\1\\\end{bmatrix}, \quad
v_2 = v_0 + e_5 = \frac{1}{2}
    \begin{bmatrix}-1\\-1\\-1\\\end{bmatrix}.
\end{align*}

Step 7: Using the lattice vectors $p_x$, $p_y$, and $p_z$, and the vertex vectors $v_0$, $v_1$, and $v_2$, the dual crystal structure is defined. 
The structure can be written in VASP format as:
\begin{verbatim}
Dual FCC
1.0             #scale
1 0 1           #p_x
1 1 0           #p_y
0 1 1           #p_z
H               #element
3               #vertex number
Cartesian
0.0 0.0 0.0     #v_0
0.5 0.5 0.5     #v_1
-0.5 -0.5 -0.5  #v_2
\end{verbatim}
Here, hydrogen atoms are placed at the dual vertices. Figure~\ref{fig:FccHcpBcc}b shows the resulting dual crystal structure, which is composed of rhombic dodecahedra that tile the space.

Step 8: Finally,  we convert the dual crystal structure to a crystal structure using CVT. By placing atoms at the centers of the rhombic dodecahedra and connecting them appropriately, we obtain the FCC crystal structure, as shown in Figure~\ref{fig:FccHcpBcc}c. In the resulting structure, $v_0$ corresponds to an octahedral center, and $v_1$ and $v_2$ correspond to tetrahedral centers.
This demonstrates that our method correctly reproduces the FCC structure from its dual graph.

We applied the same procedure to the HCP and BCC structures. 
The dual periodic graph of HCP is shown in Figure~\ref{fig:FccHcpBcc}d. The primitive unit cell of HCP contains four tetrahedral and two octahedral interstitial sites, meaning it consists of four tetrahedra and two octahedra. 
Unlike FCC, in the HCP structure, tetrahedra share faces with other tetrahedra, and octahedra share faces with other octahedra. 
The dual periodic graph contains four edges connected to each green vertex ($v_2$–$v_5$), and eight edges connected to red vertices ($v_0$ and $v_1$), corresponding to tetrahedral and octahedral face counts.

The dual periodic graph of BCC is shown in Figure~\ref{fig:FccHcpBcc}g. 
The primitive unit cell of BCC contains six tetrahedral interstitial sites. 
Unlike FCC and HCP, BCC consists solely of tetrahedra. 
The dual periodic graph features green vertices, each connected to four edges, corresponding to tetrahedral faces.

The resulting dual and crystal structures for HCP and BCC, generated by this method, are shown in Figure~\ref{fig:FccHcpBcc}e–i. 
Although detailed procedures are omitted here, the generated crystal structures match the HCP and BCC structures, respectively.
Thus, our method successfully reconstructs crystal structures from their dual periodic graphs. 
For more details on the construction process and implementation for HCP and BCC, please refer to ref.~\cite{Yokoyama2023}.

\section{Challenges and Outlook}
In this highlight, we have demonstrated that FCC, HCP, and BCC structures can each be generated from their respective dual periodic graphs using our proposed methodology. 
Our method can be applied to generate a wide range of crystal structures—including undiscovered ones.
However, our approach faces several technical challenges. 
In this section, we outline five key issues that must be addressed for practical and scalable applications of our method.

\subsection*{Combinatorial Explosion in Closed-Path Selection}
A major challenge lies in selecting an appropriate set of closed paths when the Betti number $b$ exceeds 3, which is typical for 3D periodic graphs. 
In such cases, the structure must be projected from a high-dimensional closed-path space onto 3D space, requiring a suitable selection of $b$ linearly independent closed paths. 
For complex crystal structures, the number of possible closed-path combinations increases exponentially, making the selection process extremely challenging.

For instance, the dual periodic graph of the BCC structure contains 63 unique closed paths. 
To construct the correct dual BCC structure, one must select 7 closed paths (the Betti number), such as three 3-vertex closed paths, three 4-vertex closed paths, and one 6-vertex closed path.
The total number of such combinations amounts to 407,680.
Among them, only 128 combinations yielded the correct structure with the highest symmetry corresponding to the Im$\bar{3}$m space group. 
The remaining combinations either produced structures with lower symmetry or failed to generate a valid crystal structure at all.

Although we identified a valid closed-path combination through trial and error, a general rule for optimal closed-path selection remains unknown.
Even for highly symmetric structures with a small number of atoms in the unit cell, such as BCC, the combinatorial complexity makes it difficult to guarantee successful reconstruction.
This combinatorial challenge is particularly pronounced for chemically complex systems such as MOFs or molecular crystals, which often feature intricate connectivity and large unit cells, resulting in high Betti numbers that render manual selection infeasible.
Solving this problem is essential for applying the method to more complex crystal structures. 
In future work, we aim to develop efficient algorithms to identify valid closed-path combinations. 

\subsection*{Extension to Multi-component and Low-Symmetry Structures}
While the examples presented in this work primarily focus on single-element frameworks or simple binary orderings, extending this methodology to multi-component systems (ternary or higher) is a critical future direction. 
A fundamental geometric challenge in such systems is that they inherently involve different bond lengths for different atomic pairs. 
The current standard realization formalism treats all edges equally (unit weight), which tends to produce uniform edge lengths in the generated structure. 
However, when considering pairs of different atoms, unequal edge weights should be considered. Indeed, the theory by Kotani and Sunada discusses the general case involving edge weights.\cite{Kotani2001}
In this context, weights are introduced to represent preferred interatomic distances. 
By utilizing this weighted framework, it becomes possible not only to capture the chemical diversity of multi-component systems but also to intentionally generate low-symmetry or distorted structures directly from the graph topology, thereby significantly broadening the scope of this generative approach. 

\subsection*{Integration into Materials Discovery Workflow}
To utilize this method in practical materials discovery, it should be viewed as a specialized structure generation engine within a comprehensive workflow. Our method generates idealized high-symmetry frameworks based solely on topological information, without inherent compositional data or physical scale. Therefore, a practical workflow involves the following steps:

\begin{enumerate}
\item Topology Generation: The dual periodic graph is converted into a high-symmetry crystal framework using the standard realization method.
\item Composition Optimization: Since the generated framework defines only atomic sites, specific elements must be assigned to these sites. This is a combinatorial optimization problem; for instance, determining which sites should be occupied by specific anions or cations to minimize the internal energy. Techniques such as Ising machines can be effectively applied here to find optimal atomic configurations.\cite{Ichikawa2024,Ichikawa2025}
\item Structural Relaxation: The structure with assigned elements serves as a high-quality initial guess. Subsequent geometry optimization using Density Functional Theory (DFT) or Machine Learning Interatomic Potentials (MLIP) is essential to relax lattice constants and internal coordinates to a physically realistic local minimum.
\end{enumerate}

Furthermore, these relaxed structures can be further utilized as initial populations for evolutionary algorithms or as training data for deep generative models, thereby creating a cycle of continuous structure generation and refinement.

\subsection*{Strategies for the Inverse Problem: From Property to Graph}
A central challenge in applying this framework to materials design is the inverse problem: determining which dual periodic graph yields a desired material properties. 
We propose a knowledge-driven approach to establish this property-graph relationship when crystallographic heuristics are known.

For example, based on the knowledge that tetrahedral-to-tetrahedral site transitions have low activation energies in ionic conductors,\cite{Wang2015,Yokoyama2024} 
we can generate a dual graph constrained to consist only of tetrahedra. 
Alternatively, it has been established that the mode of connectivity (corner-, edge-, or face-sharing) dictates the band gap size in optical materials.\cite{kamminga2017role} 
Similarly, for catalytic materials, the Generalized Coordination Number (GCN) formulation connects the topological arrangement of atoms to surface reactivity.\cite{calle2014fast} 
By constructing graphs that enforce specific coordination environments, we can target structures with desired properties.

To implement this approach in a fully generative context, it is necessary to rigorously define a dual periodic graph from such abstract polyhedral inputs.
This involves determining the number of dual vertices and their connections based on the faces of adjacent polyhedra. 
For instance, given the shapes of target polyhedra, one can determine which faces are compatible for joining, thereby constructing a dual periodic graph. 
Once all possible combinations of such graphs are enumerated, our method can, in principle, generate all crystal structures derivable from that polyhedral set.

\subsection*{Synergy with Data-Driven approaches}
Finally, we highlight the synergistic potential between our discrete geometric approach and modern data-driven approaches. 
As discussed in the previous section, deep generative models are promising but face significant hurdles.
Specifically, their reliance on large datasets limits their ability to extrapolate beyond known structures, and their probabilistic nature often results in structures that lack strict geometric fidelity—such as high symmetry or precise polyhedral connectivity.

Our method provides a direct solution to these challenges. 
Because it is grounded in the mathematical theory of standard realization, our approach inherently guarantees high symmetry and rigorous topological connectivity without requiring any training data. 
This distinct advantage allows our method to complement data-driven approaches in two key ways. 
First, it can generate diverse, physically valid, and highly symmetric structural templates to augment training datasets, thereby improving the data quality and extrapolative power of deep learning models. 

Second, it can serve as a deterministic geometric regularizer, transforming potentially distorted or noisy topologies proposed by probabilistic models into precise, realizable crystal frameworks. 
This hybrid strategy—combining the broad exploration capabilities of generative models with the rigorous geometric guarantees of our graph-theoretic framework—represents a robust path forward for overcoming the current limitations in inverse materials design.

\section{Conclusion}
In this highlight, we introduced a novel methodology for crystal structure generation based on polyhedral information using dual periodic graphs. By conceptualizing crystal structures as space-filling arrangements of polyhedra and constructing their corresponding dual graph representations, we developed a systematic framework grounded in discrete geometric principles.

This method was successfully applied to generate FCC, HCP, and BCC structures, demonstrating its capability to accurately reconstruct known crystal structures. The approach bridges the gap between abstract topological design and tangible crystal structure generation, offering a powerful tool for materials discovery.

Despite its potential, several challenges remain—particularly in automating the selection of closed-path bases. Nevertheless, this study represents a first step toward generating previously unknown crystal structures from target polyhedra. By enabling structure-driven materials development, our approach has the potential to identify highly functional materials that are difficult to discover through conventional composition-driven methods.



\section*{Data availability}

No primary research results, software or code have been included and no new data were generated or analysed as part of this highlight.








\bibliography{rsc} 
\bibliographystyle{rsc} 

\end{document}